# Optically reconfigurable canalization of exciton-polaritons in a non-hyperbolic perovskite

Jiahao Ren[1,#], Olha Bahrova[2,3,#], Feng Jin[1], Hao Zheng[1], Dmitry Solnyshkov[2,4], Cheng-Wei Qiu[5,*], Guillaume Malpuech[2,*], and Rui Su[1,6,*]

[1]Division of Physics and Applied Physics, School of Physical and Mathematical Sciences, Nanyang Technological University, Singapore, Singapore

[2]Institut Pascal, PHOTON-N2, Université Clermont Auvergne, CNRS, Clermont INP, F-63000 Clermont-Ferrand, France

[3]B. Verkin Institute for Low Temperature Physics and Engineering of the National Academy of Sciences of Ukraine, 47 Nauky Ave., Kharkiv 61103, Ukraine

[4]Institut Universitaire de France (IUF), 75231 Paris, France

[5]Department of Electrical and Computer Engineering, National University of Singapore, Singapore, Singapore

[6]School of Electrical and Electronic Engineering, Nanyang Technological University, Singapore, Singapore

[#]These authors contributed equally to this work.
[*]E-mails: chengwei.qiu@nus.edu.sg (C. Q.); guillaume.malpuech@uca.fr (G. M.); surui@ntu.edu.sg (R. S.)

**Abstract**

The ability to steer polariton flow on-demand holds significant promise towards nanophotonic applications and photonic circuitry. Polariton canalization, exhibiting intrinsic collimation and diffractionless transport, emerges as a promising solution without guiding structures. However, earlier demonstrations have been restricted to certain crystal surfaces with intrinsic hyperbolic responses and operated in the linear regime. Here, we experimentally demonstrate canalization of nonlinear exciton-polariton condensates with optical reconfigurability in a birefringent $CsPbBr_3$ perovskite crystal without intrinsic hyperbolic response. By embedding the birefringent perovskite crystal into a planar microcavity, the interplay between cavity transverse-electric–transverse-magnetic splitting and crystalline birefringence produces an anisotropic band geometry with a hyperbolic–flat–parabolic evolution of polaritonic isofrequency contours (IFCs). Nonresonant pumping drives exciton polariton condensation onto flat far-field contours with nonlinear emission amplification, leading to coherent canalized flows with over twentyfold collimation with respect to arc-shaped contours. Reconfiguring the optical pumping-spot size allows switching the nonlinear polariton condensates into hyperbolic and parabolic IFC regimes, leading to divergent propagation behaviour with collimating reconfiguration. Our study reveals a distinct canalization framework for shaping the nonlinear exciton-polariton condensate flows, opening opportunities for all-optical polaritonic logic circuits based on stabilized nonlinear quantum interconnects.

**Main**

Diffraction inherently induces transverse spreading and loss in the propagation of all types of waves and quantum particles, positioning diffraction-free canalization as a long-standing target for efficient energy transfer[1–12]. Canalization in photonics, where light travels in a self-guided, highly directional manner without the need for designed waveguide channels, provides a compelling approach for energy-efficient photonic routing and manipulation[2–12]. Polaritons, hybrid light-matter quasiparticles resulting from the strong coupling between photons and matter excitations, have recently emerged as a powerful platform for achieving canalized propagation at the nanoscale[3–15]. Their promise lies in the capability to precisely tune the polaritonic isofrequency contours (IFCs), which are constant-frequency slices of the polaritonic dispersion in momentum space that dictate the directionality of polariton flow[3–12,16]. Nevertheless, existing realizations of polariton canalization have relied heavily on the intrinsic hyperbolic response of certain natural crystal interfaces for strongly anisotropic dispersion tailoring toward flat IFCs[3–12], including metasurfaces[4], twisting[3,5,7] and heterostructuring[11] applied to surface phonon polaritons, as well as ion intercalation for surface plasmon polaritons[8]. The restricted applicability of polariton canalization substantially challenges their broader scalability and potential applications. Moreover, earlier demonstrations with surface polaritons were usually limited to the linear regime, where the output scales linearly with the input. Further access to coherent nonlinear regimes, such as optical amplification with reconfigurability, is highly desirable but remains challenging.

Microcavity exciton-polaritons, resulting from the superposition of cavity photons and semiconductor excitons in the strong coupling regime[17,18], provide a distinctive route toward tunable nonlinear condensates with spontaneous bosonic coherence for canalized polaritons. As composite bosonic quasiparticles, exciton-polaritons can undergo nonlinear polariton condensation through bosonic stimulation at elevated temperatures[19–23], which gives rise to a range of collective quantum phenomena[24], such as superfluidity[25], quantized vortices[26] and rich polarization textures[27]. In particular, their inherently strong polariton interactions introduce pronounced optical nonlinearity[28], enabling gain-induced emission amplification[29–31] and on-demand reconfigurability of polariton condensation behaviour in energy, momentum, and polarization by modulating the optical excitation beam[32–36]. Despite these exciting prospects, nonlinear exciton-polariton condensates were mostly limited to the isotropic band geometry, offering little scope for dispersion engineering toward canalization[33,37,38]. This motivates the formulation of an alternative route to canalized polaritons, circumventing the stringent requirement of intrinsic material hyperbolicity for anisotropic band shaping.

In this work, we demonstrate a distinct regime of optically reconfigurable canalization of nonlinear exciton-polariton condensates with a birefringent $CsPbBr_3$ perovskite crystal lacking intrinsic hyperbolic behavior. By embedding the birefringent $CsPbBr_3$ perovskite into a planar microcavity, the splitting of the transverse-electric (TE) and transverse-magnetic (TM) modes in the optical microcavity interacts with the intrinsic birefringence of the perovskite, giving rise to an anisotropic band geometry in which the two polarization branches intersect at degenerate tilted Dirac points[39,40]. Above the tilted Dirac points, the polaritonic IFCs exhibit a hyperbolic–flat–parabolic evolution in the far field. Exciton-polaritons can be driven under appropriate nonresonant pulsed excitation to condense onto the flat IFCs with superlinear emission amplification, producing highly collimated coherent polariton flows in the canalization regime. By adjusting the optical pumping spot size, the nonlinear polariton condensates can be shifted away from the flat IFCs into hyperbolic or parabolic regimes, giving rise to real-space propagation with increasing divergence. Quantitative analysis of the lateral broadening dynamics across different regimes reveals a reconfigurable enhancement in polariton condensate collimation by 3.4 to 20.5 times over the arc-shaped IFC–induced divergence, opening new opportunities for all-optical nonlinear polaritonic devices with controllable directional confinement.

**Results**

**Anisotropic IFC evolution in a birefringent perovskite microcavity**

To create a novel canalization mechanism through anisotropic exciton-polariton band geometry that induces IFC curvature evolution, we employ the halide perovskite $CsPbBr_3$ with an orthorhombic structure[30], which is intrinsically birefringent and optically anisotropic along distinct crystalline axes, as the active medium in a distributed Bragg reflector (DBR) cavity (Fig. 1a). The system exhibits both TE–TM splitting[41] and constant linear birefringence, forming a polarization-mode doublet described in a circular polarization basis by an effective Hamiltonian:

$$H_k = \begin{pmatrix} E_0 + \frac{\hbar^2 k^2}{2m} & \beta_0 - \beta k^2 e^{-2i\varphi} \\ \beta_0 - \beta k^2 e^{2i\varphi} & E_0 + \frac{\hbar^2 k^2}{2m} \end{pmatrix}, \quad (1)$$

where $m = m_{TM}m_{TE}/(m_{TM} + m_{TE})$, $k = |\mathbf{k}| = \sqrt{k_x^2 + k_y^2}$ is the in-plane wavevector ($k_x = k\cos\varphi$, $k_y = k\sin\varphi$), $E_0$ is the mode energy at $k = 0$, $\beta$ is the strength of TE-TM splitting, and $\beta_0$ represents the optical birefringence from the perovskite. Diagonalization of this effective Hamiltonian yields two eigenenergies, associated with orthogonal linear polarizations of the photonic modes

$$E_{ph\pm} = E_0 + \frac{\hbar^2 k^2}{2m} \pm \sqrt{\beta_0^2 + \beta^2 k^4 - 2\beta_0 \beta k^2 \cos 2\varphi}. \quad (2)$$

The exciton-polariton dispersion is quite accurately captured by a coupled exciton–photon model incorporating the energy dependence of the exciton-polariton effective mass under a Rabi splitting $\Omega_R$ and neglecting the energy difference between the two orthogonal exciton polarizations at the exciton resonance ($E_{ex}$), giving the lower and upper polariton branches

$$E_{pol} = \frac{E_{ex} + E_{ph\pm}}{2} \pm \frac{1}{2}\sqrt{[E_{ph\pm} - E_{ex}]^2 + \Omega_R^2}. \quad (3)$$

To reveal the anisotropic exciton-polariton dispersion, Fig. 1b presents three-dimensional band-view maps of the $S_1$ Stokes parameter for the experimentally measurable lower branches, calculated from equations (1)-(3), with $y$-polarized contours plotted at iso-energy intervals. At a critical frequency the two polarization eigenmodes form two shifted circles which intersect at the tilted Dirac point coordinates $k_{x,Dirac} = \pm(\beta_0/\beta)^{1/2}$ where the two polarization modes are degenerate. This singularity occurs because of the exact balance between the TE–TM splitting and linear birefringence. Above the tilted Dirac points, the diffraction of a $y$-polarized wave-packet propagating along $k_x$ can be evaluated by computing the inverse of the transverse effective mass (that is, band curvature) which reads

$$\left.\frac{\partial^2 E}{\partial k_y^2}\right|_{(k_y=0)} \simeq \underbrace{\frac{\hbar^2}{m}}_{\text{isotropic curvature}} - \underbrace{\frac{4\beta_0\beta}{|\beta_0-\beta k_x^2|}}_{\text{anisotropic curvature}} \quad (4)$$

This expression contains the isotropic diffraction term $\hbar^2/m$ induced by a circular IFC and an anisotropic contribution which induces divergence near the tilted Dirac point. For $k_x$ slightly larger than $k_{x,Dirac}$ the solution with the minus shows a negative curvature corresponding to a hyperbolic IFC (blue in Figure 1.b). This curvature vanishes exactly at $k_{x,Flat} = k_{x,Dirac}\sqrt{1+\frac{4m}{\hbar^2}\beta}$ yielding a flat IFC (red in Figure 1.b) that marks the onset of canalization. With further increase of the energy, $|\beta_0 - \beta k_x^2|$ grows, the anisotropic contribution diminishes, and the dispersion gradually returns to a parabolic profile dominated by the isotropic curvature (green in Figure 1.b). Under strong light-matter coupling, the same curvature-compensation condition remains valid, with parameters slightly renormalized by the photonic Hopfield coefficient $C$, which reflects the photon fraction of the hybrid mode.

When viewed through polarization selection, the IFC evolution reveals a pronounced angular anisotropy. The polarization of the modes is obtained by computing the eigenvectors of the effective Hamiltonian equation (1). Above the energy of tilted Dirac points, the *y*-polarized branch defined by the modes with $S_1 = (I_x - I_y)/(I_x + I_y) < 0$ persists only within a narrow angular sector around $\varphi \approx 0$ along $k_x$, as illustrated in Fig. 1b. This restriction causes the *y*-polarized IFC to split into two non-closed segments directed toward $\pm k_x$, which funnel the group-velocity vectors into a narrow cone and establish the origin of the enhanced directional polariton flow along $x$. A magnified view of the positive-$k_x$ region in Fig. 1b reveals that, as energy increases, *y*-polarized IFCs with isolated segments evolve successively from hyperbolic to flat and parabolic geometries. The corresponding real-space propagation is captured by schematic wavefronts reconstructed from the two-dimensional (2D) momentum-space contours in Fig. 1c, which evolve from hyperbolic divergence to flat collimation (canalization) and finally to parabolic divergence. In comparison, toward $\pm k_y$, the IFCs remain strongly curved across broad angular ranges, forming a typical arc-shaped regime in sharp contrast to the confined segments toward $\pm k_x$. These isolated IFCs toward $\pm k_x$ with distinct geometries thus provide the possible reconfigurability of the wavefront morphology and directional transport by changing the frequency of the propagating beam, with the flat contours yielding the highest degree of collimation.

**Canalization of nonlinear exciton-polariton condensates**

To achieve optical reconfigurability of nonlinear exciton-polariton condensates around the canalization regime, we employ a flexible yet elegant spot-size tuning scheme under nonresonant pulsed excitation, allowing active control of IFC configurations at specific energies and momenta. Such excitation spot-size control has previously been shown to provide the access to different condensation energies and momenta in other nonlinear exciton-polariton systems[32,36]. Nonresonant pumping of the system creates a local exciton reservoir that interacts strongly and repulsively, generating a spatially confined and strongly blueshifted polariton condensate region that acts as an effective potential hill[33,42].

The pump-induced polariton condensate initially accumulates at the top of this potential hill, converting potential energy into kinetic energy and thereby acquiring a finite in-plane wave vector[29]. When the excitation spot is far smaller than the polariton propagation length, most polaritons escape the potential hill and propagate ballistically away from the pumped region[32,43]. In this regime, further reducing the pump-spot size concentrates the local exciton reservoir, thereby raising and steepening the potential hill that places the condensate at higher energies and in-plane momenta, which enables the optical reconfiguration of the polariton condensation energy and momenta. To realize polariton condensation in the canalization regime, the birefringent perovskite microcavity is excited nonresonantly by 400 nm pulsed laser with a 1.6 μm diameter pump spot. Below the condensation threshold ($P < P_{th}$), we experimentally demonstrate the establishment of anisotropic polariton dispersion, where two orthogonally linear-polarized branches along $k_x$ at $k_y = 0$ intersect at a pair of tilted Dirac points at 2.320 eV (with $k_x \approx \pm 4.00$ μm$^{-1}$), as measured by angle-resolved photoluminescence spectroscopy (Fig. 2a, left panel; Supplementary Fig. S2b). When the excitation power is increased to a fluence of 29.4 μJ cm$^{-2}$ ($P = 4P_{th}$), massive accumulation of polaritons on the $y$-polarized branch leads to pronounced emission feature at $k_x \approx 4.65$ μm$^{-1}$ and 2.326 eV, suggesting the occurrence of polariton condensation, as shown in the right panel of Fig. 2a. The 2D momentum-space image in Fig. 2b reveals that the polariton condensate at $k_x \approx 4.65$ μm$^{-1}$ appears on a flat IFC segment, with polaritons concentrated and isolated toward $+k_x$ in the $y$-polarized emission. To evaluate the polarization dependence of the flat-IFC condensate, we extracted the emission intensity at $k_x \approx 4.65$ μm$^{-1}$ from the dispersion along $k_x$ at $k_y = 0$ as a function of the detection polarization angle (Supplementary Fig. S3), revealing a high linear polarization degree exceeding 90% along the perovskite crystal $y$ axis. To further demonstrate the polariton condensation behaviour associated with the flat IFC, we traced the nonlinear phase transition by quantitatively extracting the evolution of the integrated emission intensity, linewidth, and peak position of the dispersion interval near $k_x = 4.65$ μm$^{-1}$ as a function of increasing pumping fluence, as shown in Fig. 2c. Upon crossing the threshold ($P_{th} = 7.4$ μJ/cm²), polaritons condense near 2.326 eV, with emission intensity undergoing nonlinear amplification by four orders of magnitude and linewidth collapsing to 2 meV, while the repulsive polariton interaction causes a continuous blueshift as the pump fluence increases.

The polariton canalization of the flat-IFC condensate is further revealed by real-space emission measurements as a collimated propagation along the $x$ direction, as shown in Fig. 2d. By normalizing the intensity within a 30° sector toward $\pm k_x$ along $y$ for each value of $x$ coordinate, the essentially constant width of the polariton wave packet during propagation becomes clearly visible. This highly directional polariton condensate propagation toward $\pm x$ arises from bilaterally flat contours near $k_x = \pm 4.65$ μm$^{-1}$ in 2D momentum space (Supplementary Fig. S4), which induce collinear Poynting vectors perpendicular to the contours, thereby defining a common polariton group velocity direction and suppressing divergence. In sharp contrast, polaritons propagating along the $\pm y$ directions exhibit conventional spatial diffusion, owing to the typical arc-shaped features of the IFC directed toward $\pm k_y$. To circumvent the substantial pump-induced perturbations to the polariton condensate, including the effects of hole burning[36,44,45] and real-space collapse[46], we focus on the polariton flow downstream along +x, starting 5 μm from the excitation spot, and normalizing the intensity along y at each x coordinate to exclude dissipative contributions when assessing the canalized flow over a 20 μm region (Fig. 2e). The persistence of the polariton wave-packet width over long distances highlights the robustness of the canalized propagation. This observed canalization is well supported by the simulation results (Fig. 2e). Although the canalized polariton condensate does not occupy the ground state of the system, it can still be regarded as a nonequilibrium coherent condensate that inherits coherence from the condensation[29]. To assess this coherence, we use a Mach–Zehnder interferometer to overlap the real-space emission of the canalized condensate with its mirror image, which produces clear interference fringes along the canalization channel (Supplementary Fig. S6) and confirms that spatial coherence is preserved during propagation.

**Collimation evolution across the IFC transition**

The flat IFC that drives canalization delineates the transitional frontier between hyperbolic and parabolic dispersions. To trace the pronounced collimation effect through the transition, we experimentally realized polariton condensation at energies slightly below and above the flat-IFC position, thereby capturing the condensates on either side of the canalization phase. Figures 3a and 3e (right panels) show condensates at two additional energy positions in the dispersion, namely 2.323 eV (at $k_x \approx 4.30$ μm$^{-1}$, hyperbolic) and 2.334 eV (at $k_x \approx 5.55$ μm$^{-1}$, parabolic), which are realized *in situ* on the same sample, while maintaining constant pump power, by tuning the excitation spot diameters to 2.0 μm and 0.8 μm, respectively (details in Methods). The corresponding *S1* Stokes parameter maps of the dispersions reveal the high-purity *y*-polarization associated with condensates propagating along the *x* direction (left panels of Figs. 3a and 3e). At 2.323 eV, the polariton condensates exhibit the 2D momentum space emission with hyperbolic segments at $+k_x$ (Fig. 3b), whose two slightly curved lobes generate non-collinear, inward-pointing Poynting vectors, thereby shaping the wavefront propagation. Therefore, outside the disturbance-dominated near-pump region, the observed hyperbolic polariton flow exhibits a slight divergence along $+x$, resulting in a weak gradual broadening of the wave packet over a 20 μm downstream region (Fig. 3c), consistent with the theoretical results in Fig. 3d. When condensation occurs at 2.334 eV in the parabolic regime (Fig. 3e), the IFC reshapes into $+k_x$-oriented parabolic segments in 2D momentum space emission (Fig. 3f), whose parabolic geometry generates non-collinear, outward-pointing Poynting vectors, resulting in a diverging outward wavefront. Figure 3g shows a divergence of the real space polariton flow downstream along $+x$ in the parabolic regime, forming a substantially broadened wave packet compared with the hyperbolic and canalized cases, consistent with the theoretically predicted divergent evolution in Fig. 3h.

To quantitatively assess the governing role of IFC morphology in directional polariton flow, we extracted the evolution of the real-space wave-packet width along *x* for the hyperbolic, flat, and parabolic regimes, and compared it with the strong *y*-axis divergence driven by the arc-shaped contours at $-k_y$ (Fig. 4a). Linear fitting of the polariton beam-width evolution over the propagation distance yields lateral broadening rates ($\Gamma_{\text{regime}}$) of 0.102 ± 0.020 ($\Gamma_{\text{hyperbolic}}$), 0.064 ± 0.007 ($\Gamma_{\text{flat}}$), 0.381 ± 0.035 ($\Gamma_{\text{parabolic}}$), and 1.309 ± 0.060 ($\Gamma_{\text{arc-shaped}}$), respectively. To demonstrate the dynamic reconfigurability of polariton condensate flow, we further realized experiments with intermediate IFC (between flat and parabolic) that we call weak-parabolic states, which demonstrates our capacities of continuous tuning of the canalization effect. These experiments are reported in Supplementary Figs. 7-9 and correspond to condensation energies of 2.328 eV (at $k_x \approx 4.86$ μm$^{-1}$), 2.330 eV (at $k_x \approx 5.08$ μm$^{-1}$), and 2.332 eV (at $k_x \approx 5.31$ μm$^{-1}$). The quantitative analyses of their wave-packet evolution along $x$ yield lateral broadening rates of 0.076 ± 0.011, 0.111 ± 0.015, and 0.270 ± 0.021. Accordingly, using the strongly divergent arc-shaped case as a benchmark ($\Gamma_{\text{arc-shaped}}$), we define the collimation factor of each regime as $\Gamma_{\text{arc-shaped}}/\Gamma_{\text{regime}}$ to quantitatively track the evolution of directional collimation across the IFC transition (Fig. 4b). The collimation factor is expressed as a function of the normalized energy offset $\xi = (E - E_D)/(E_c - E_D)$, where $E$ is the condensation energy for each operating point, $E_D$ denotes the tilted-Dirac-point energy, and $E_c$ corresponds to the canalization energy. It peaks near canalization at $\xi \approx 1$ with a value of 20.5 ± 2.4. On the hyperbolic side with $\xi < 1$ it decreases but remains high, reaching 12.8 ± 0.6 at $\xi \approx 0.5$. For even smaller $\xi$ approaching 0, the $y$-polarized IFC segments with the group velocities directed toward $+k_x$ fade as illustrated in Fig. 1d, which nearly suppresses transport along +x and renders the flow unobservable. On the parabolic side with $\xi > 1$, the collimation factor declines progressively until it reaches the experimental bound of 3.4 ± 0.4 at $\xi \approx 2.3$. The black dashed curve in Fig. 4b represents the theoretical prediction for collimation-factor evolution, which captures both the canalization peak and the asymmetric downward trend on either side. Overall, the evolution exhibits continuous tunability of directional polariton collimation over a broad range approaching one order of magnitude, achieved simply by shifting the condensation energy across the hyperbolic-flat-parabolic sequence.

**Discussion**

In summary, we demonstrate exciton-polariton canalization featuring optically reconfigurable nonlinear condensation in a microcavity utilizing a birefringent perovskite crystal without intrinsic hyperbolicity. Remarkably, the collimation of canalized polariton condensate exhibits over twenty times improvement relative to the typical divergence for arc-shaped IFCs, while the IFC selection controlled by the optical pump-spot size affords nearly an order of magnitude of tunability in directional confinement. Our results not only establish a novel paradigm for optical manipulation of polariton condensates in structure-free configurations that enables diffraction-negligible canalization, but also enhance the sensitivity and agility of the on-demand canalization mechanism.

In contrast to previous demonstrations of surface polariton canalization limited to certain crystal surfaces with intrinsic hyperbolic responses[6,12], we lift this critical requirement and demonstrate a general mechanism to access the reconfigurable polariton canalization regime, which in principle applies to any birefringent crystal. Moreover, the canalized exciton-polariton condensates provide a controllable gateway to nonlinear regimes, featuring superlinear emission amplification[29–31] and offering an ideal platform for interacting quantum fluids that naturally foster further steering of polariton flows, including solitons[47,48], spin-current focusing[49], and stripe modulations[39]. Further, the reconfigurable nonlinear exciton-polariton condensates with spontaneous coherence, as a macroscopically single quantum state, promise significant advances in efficient quantum-coherent information processing through canalized routing[50], which opens a bright future for programmable polaritonic logic[51,52] in quantum networks[53,54] and neuromorphic computing[55,56].

## Methods

### Sample fabrication

The perovskite planar microcavity consists of a bottom distributed Bragg reflector (DBR), a polymethyl methacrylate (PMMA) spacer, a birefringent all-inorganic perovskite (CsPbBr$_3$) platelet, a ZEP520A spacer, and a top DBR. In fabrication, 15.5 pairs of alternating tantalum pentoxide (Ta$_2$O$_5$) and silicon dioxide (SiO$_2$) layers are deposited onto a silicon substrate using an electron-beam evaporator (Cello 50D) to form the bottom DBR. Following plasma surface treatment, a 30-nm-thick PMMA spacer is spin-coated onto the bottom DBR. High-quality single-crystalline CsPbBr$_3$ platelets (~300 nm thick) are synthesized via chemical vapor deposition (CVD), where a quartz crucible containing CsPbBr$_3$ powder and a 100 μm-thick mica substrate are placed at the centre and downstream zone of the CVD furnace, respectively. The chamber pressure is maintained at 43 Torr with a 30 standard cubic centimetres per minute (sccm) nitrogen flow (99.999% purity). The CVD furnace temperature is ramped up to 590 °C within 5 min, held for 10 min, and then naturally cooled to room temperature, resulting in perovskite platelet growth on the mica substrate[31,57]. The platelets are transferred onto the PMMA-coated DBR using a dry tape-transfer technique, while the mica substrate is removed with Scotch tape[58]. Finally, a 50 nm ZEP520A layer is spin-coated as a protective spacer, followed by electron-beam deposition of 8.5 pairs of Ta$_2$O$_5$/SiO$_2$ layers as the top DBR to complete the microcavity structure.

### Optical spectroscopy characterizations

Optical characterization of the perovskite planar microcavity is carried out using a home-built angle-resolved spectroscopy setup equipped with Fourier optics, operated in a reflection configuration under non-resonant pulsed excitation at 400 nm (Supplementary Fig. S1). The 400 nm pulsed pump is generated by second-harmonic generation (SHG) through a β-BaB$_2$O$_4$ (BBO) crystal from the 800 nm output of a mode-locked Ti:sapphire oscillator (Spectra-Physics; repetition rate: 80 MHz; pulse duration: 100 fs). Prior to entering the optical setup, the pulsed beam is precisely conditioned by a *4f* optical system with a pinhole spatial filter, allowing arbitrary beam shaping and its subsequent normal incidence onto the sample surface as a Gaussian spot with a finely tunable diameter ranging from 1.0 to 2.0 μm. Specifically, with the pump power kept essentially constant, the excitation spot size is tuned to ~2.0 μm, ~1.6 μm, ~1.4 μm, ~1.2 μm, ~1.0 μm, ~0.8 μm, which correspond to pump fluences of 18.8 μJ cm$^{-2}$, 29.4 μJ cm$^{-2}$, 38.4 μJ cm$^{-2}$, 52.3 μJ cm$^{-2}$, 75.3 μJ cm$^{-2}$, 117.6 μJ cm$^{-2}$, respectively. This spot-size control enables *in situ* tuning of the polariton condensation energy to 2.323 eV, 2.326 eV, 2.328 eV, 2.330 eV, 2.332 eV, and 2.334 eV, respectively, as demonstrated in the main text. The resulting spot-size–induced modulation of the condensation energy is significantly stronger than the density-induced variation shown in Fig. 2c, a consequence of the distinct intrinsic mechanism[36].

During the optical measurements, the sample is mounted in a closed-cycle cryostat (Montana

Instruments) equipped with a piezo-controlled three-dimensional translation stage, vacuum housing, and radiation shield, maintaining a stable environment at liquid-nitrogen temperature (77 K) to prevent thermal damage caused by the tightly focused, high-intensity pulsed excitation, even though the perovskite device is known to operate well at room temperature from numerous previous studies[22,59,60]. A window-corrected objective (Olympus LCPlan N 50X, NA 0.65, WD 4.5 mm) outside the cryostat focuses the pulsed beam onto the sample and collects the photoluminescence emission. For momentum-space emission measurements, a *4f* imaging system projects the back focal plane of the objective onto the entrance slit of a spectrometer (Horiba iHR550, 550 mm focal length) equipped with a 600 lines/mm grating that provides energy-resolving capability, and a liquid-nitrogen-cooled charge-coupled device (CCD, 256 × 1024 pixels). For real-space emission measurements, the real plane is directly imaged onto the CCD detector, while interference fringe images are obtained by splitting and recombining the real-space emission in a Mach–Zehnder interferometer (inset of Supplementary Fig. S1). For linear polarization analysis, a half-wave plate (HWP) and a linear polarizer are sequentially inserted into the detection path, enabling selective detection of *x*- ($I_x$) and *y*- ($I_y$) polarized photoluminescence intensities.

**Theoretical modelling of collimated polariton propagation**

In order to theoretically reproduce the different regimes of the polariton propagation observed in the experiment, we perform numerical simulations of the following spinor Schrödinger equations for the coupled photon $\psi_\pm$ and exciton $\chi_\pm$ mean-field wave functions:

$$i\hbar \frac{\partial \psi_\pm}{\partial t} = \left[-\frac{\hbar^2}{2m_{PH}}\Delta - i\frac{\hbar}{2}\gamma + U\right]\psi_\pm + \beta(\partial_y \pm i\partial_x)^2 \psi_\mp + 2\beta_0 \psi_\mp + \frac{1}{2}\Omega_R \chi_\pm + P(x,y)e^{-i\omega_p t} \quad (5)$$

$$i\hbar \frac{\partial \chi_\pm}{\partial t} = \left[U - i\frac{\hbar}{2}\gamma_{EX}\right]\chi_\pm + \frac{1}{2}\Omega_R \psi_\pm \quad (6)$$

Here, $m_{PH}$ is the photon mass (the exciton mass is considered as infinite), the coefficients $\beta$ and $\beta_0$ are the strengths of the TE-TM splitting and the linear birefringence (see also Hamiltonian equation (1) in the main text). We take $2\beta_0 = 18$ meV from a fit of data of the measured dispersion (see Supplementary Figure 10). The strong anisotropy of the dispersion is accounted for by the photon mass anisotropy: $\beta = \hbar^2/4 * (1/m_{PH,l} - 1/m_{PH,t})$, $m_{PH,l}/m_{PH,t} = 0.665$; and $m_{PH,t} = m_{PH} = 2.25 * 10^{-5} m_0$, where $m_0$ is the free electron mass. Further, $\Omega_R = 104$ meV is the Rabi splitting that we also obtain from the fitting of the dispersion. The coefficients $\gamma$ and $\gamma_{EX}$ are the photon and exciton decay rates, respectively. The photonic decay rate is extracted from the linewidth of the experimentally-measured dispersion as $\hbar\gamma = 2$ meV and the excitonic decay rate is taken as the inverse of the exciton lifetime of 4 ps. The effect of the polariton acceleration due to the polariton-exciton scattering is taken into account by the reservoir potential $U$. As a consequence of high kinetic polariton energies, we concentrate our attention on linear regime and neglect the effects of polariton-polariton interactions. We used quasi-resonant pumping with a narrow Gaussian spot $P(x,y)$ with the frequency $\omega_p$ in order to control precisely the excitation conditions. Laplacian operators $\Delta$ were treated by Fourier transform which allows to properly access the high wave vector states. The time integration was done numerically using 3rd-order Adams-Bashforth multistep method implemented with pytorch CUDA.

The results of the numerical simulations with an appropriate treating of the experimentally accessible dispersion for three different regimes discussed thought the main text are presented in Fig. 2e, Fig. 3 (d,h), respectively. Furthermore, by tuning the condensate energy $\hbar\omega_p$, we track the evolution of the beam width along the $x$ direction and plot the resulting collimation factor in Fig. 4b (black dashed curve), which shows a good agreement with the experimental data.

**Acknowledgements**

R.S. gratefully acknowledge funding support from the Singapore Ministry of Education via the AcRF Tier 2 grant (MOE-T2EP50222-0008), AcRF Tier 3 grant (MOE-MOET32023-0003) "Quantum Geometric Advantage" and Tier 1 grant (RG90/25). R.S. also gratefully acknowledges funding supports from Nanyang Technological University via a Nanyang Assistant Professorship start-up grant. Additional support was provided by the ANR program "Investissements d'Avenir" through the IDEX-ISITE initiative 16-IDEX-0001 (CAP 20-25), the ANR project MoirePlusPlus (ANR-23-CE09-0033), and the ANR project "NEWAVE" (ANR-21-CE24-0019). We are grateful to the Mésocentre Clermont-Auvergne of the Université Clermont Auvergne for providing help, computing and storage resources.

**Author contributions**

R.S. and J.H.R. designed the research. J.H.R. synthesized the perovskite materials. J.H.R. fabricated the samples and performed all the optical measurements with the help of F.J.. O.B., G.M. and D.S. performed the theoretical calculations. R.S., J.H.R., H. Z., O.B., G.M., D.S. and C.Q. analyzed the data. R.S., J.H.R., O.B., G.M., and D.S. wrote the manuscript with the inputs from all the authors.

**Competing interests**

The authors declare that they don't have competing interests.

**Data availability**

All experimental data that support the plots within this paper are available from the corresponding author upon request.

**Code availability**

The codes are available from the corresponding author upon request.

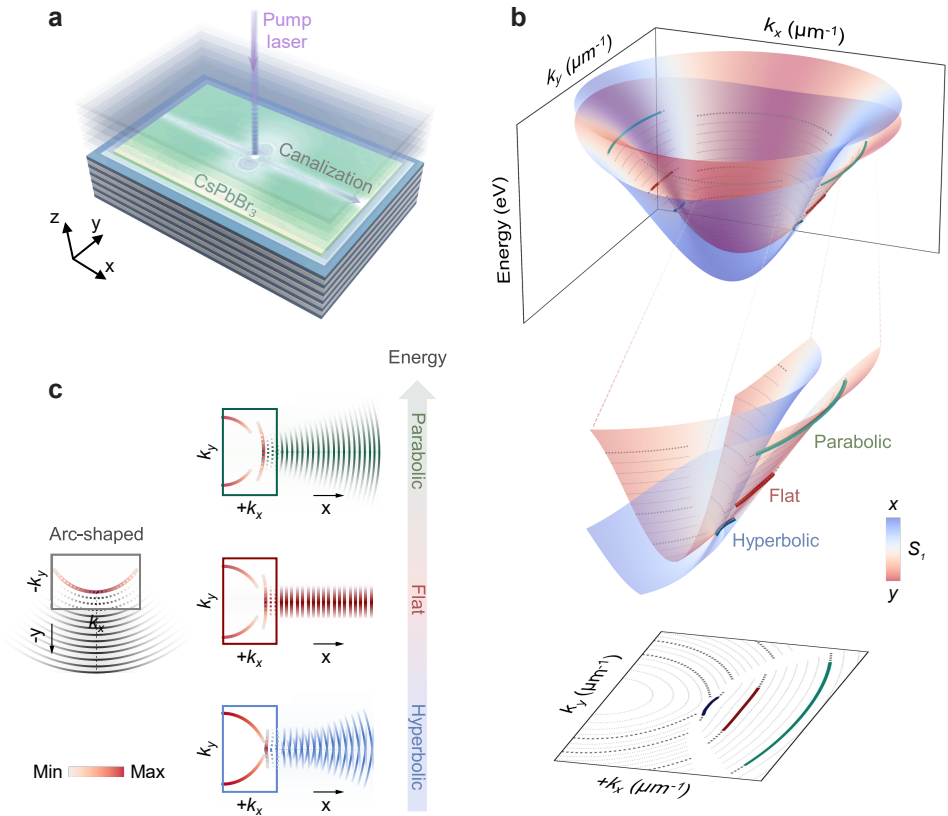

**Fig. 1| Schematic of the anisotropic band geometry in the CsPbBr₃ planar microcavity. a,** Schematic of the CsPbBr₃ planar microcavity for canalized exciton-polariton condensates. **b,** Three-dimensional band-view maps of the $S_1$ Stokes parameter for the perovskite microcavity exciton polaritons, with blue and red indicating the *x*- and *y*-polarized branches, respectively, and grey dashed lines showing *y*-polarized contours plotted at iso-energy intervals. The middle panel is magnified three-dimensional view of the positive-$k_x$ region and the bottom panel is corresponding 2D contour map of *y*-polarized IFCs. Three distinct IFC segments are highlighted in blue, red, and green, corresponding respectively to the hyperbolic, flat, and parabolic geometries above the energy of tilted Dirac points. **c,** Theoretically simulated 2D momentum-space emission (directed towards $+x$) of *y*-polarized polaritons exhibiting hyperbolic, flat, and parabolic IFCs, together with simplified real-space schematics illustrating wavefront propagation along $+x$ in blue, red, and green, respectively. These schematics respectively display hyperbolic divergence, flat collimation, and parabolic divergence, tracing the evolution of directional transport across the IFC transition and standing in stark contrast to the strongly divergent propagation along $-y$, shown in grey, induced by the typical arc-shaped IFCs.

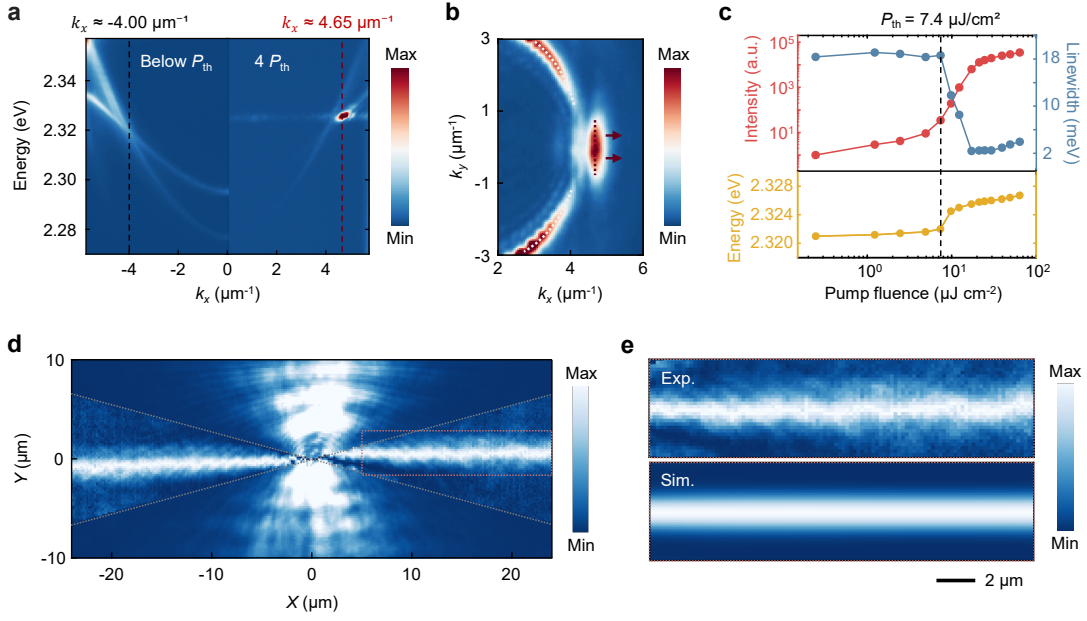

**Fig. 2| Canalized polariton flow at flat IFC condensation. a,** Angle-resolved photoluminescence spectra showing the polariton dispersion below threshold in left panel and exciton-polariton condensation (2.326 eV) on the flat IFC above threshold ($P = 4P_{th}$) in right panel. The black dashed line marks the tilted Dirac-point momentum at $k_x \approx 4.00$ μm$^{-1}$, while the red dashed line indicates the momentum of the flat-IFC condensate at $k_x \approx 4.65$ μm$^{-1}$. **b,** The 2D momentum-space image of $y$-polarized polaritons condensates on the flat IFC, showing a concentrated flat contour with the associated group velocities directed toward $\pm k_x$, highlighted by the red dotted line. The red arrows denote the directions of the Poynting vectors perpendicular to the contour, revealing the parallel and collimated polariton group velocity. The white dotted lines outline the segment belonging to the typical arc-shaped IFC at $\pm k_y$. **c,** Evolution of the integrated emission intensity, linewidth, and peak position within the dispersion interval $4.65 \pm 0.65$ μm$^{-1}$ as a function of increasing pump fluence during condensation at the flat IFC. At the threshold $P_{th} = 7.4$ μJ/cm², nonlinear amplification of the emission intensity together with linewidth narrowing strongly supports the onset of polariton condensation, while the repulsive interaction of polaritons results in a continuous blueshift. **d,** Real-space emission of canalized polariton flow induced by the flat IFC condensation under $y$-polarization, where the intensity within the 30° sector toward $\pm k_x$, delineated by the grey dotted lines, is normalized along $y$ for each $x$ coordinate to eliminate losses during polariton propagation along $x$. **e,** Magnified experimental real-space image in the top panel taken from the region outlined by the red dotted lines in (**d**), revealing canalized polariton flow toward the $+x$ direction with stable nondiffractive propagation over ~20 μm (normalized along $y$ for each $x$ coordinate). The bottom panel shows the corresponding theoretical simulation.

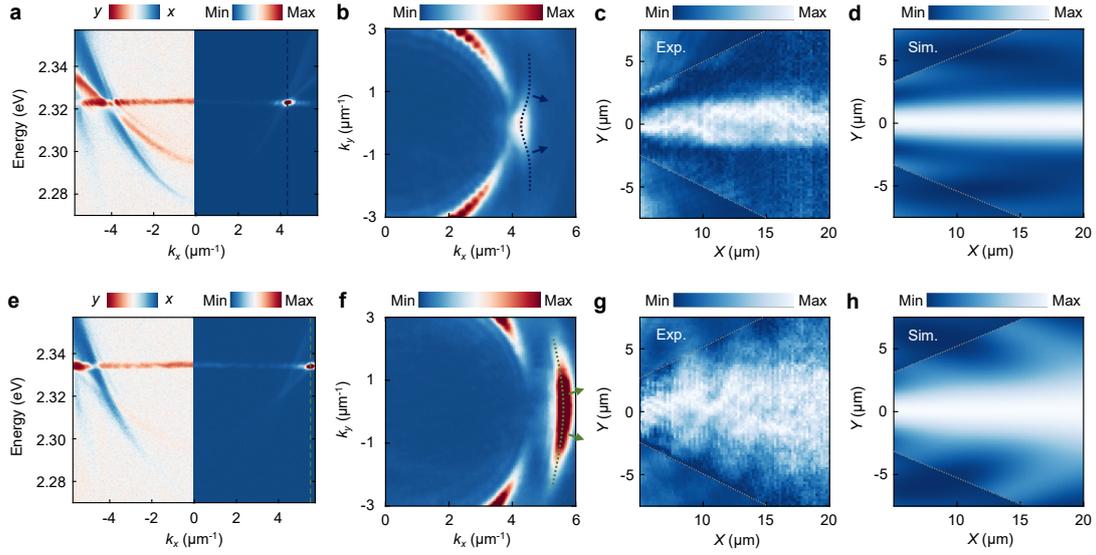

**Fig. 3| Polariton condensates in the hyperbolic and parabolic regimes. a** and **e,** Dispersions (right panel) and corresponding $S_1$ Stokes parameter maps (left panel) of polariton condensation at the hyperbolic IFC near the energy 2.323 eV (**a**) and at parabolic IFC near the energy 2.334 eV (**e**) for the same sample. These maps reveal the polarization orthogonality of two branches and demonstrate high-purity *y*-polarization associated with condensate propagation along the *x* direction. The blue and green dashed lines indicate the momenta of the hyperbolic and parabolic IFC condensates at $k_x \approx 4.30$ and 5.55 $\mu m^{-1}$, respectively. **b** and **f,** 2D momentum-space emissions under *y*-polarization at the hyperbolic IFC (**b**) and parabolic IFC (**f**) of the positive-$k_x$ region. The blue and green dotted lines mark the hyperbolic and parabolic IFC segments, and arrows of matching colour indicate the corresponding Poynting-vector directions, revealing distinct divergence behaviour of the polariton group velocity. **c** and **d,** Experimental real-space image (**c**) and theoretically simulated real-space intensity (**d**) of polariton condensates at the hyperbolic IFC, exhibiting slight divergence. **g** and **h,** Experimental real-space image (**g**) and theoretical simulation result (**h**) of polariton condensates at the parabolic IFC, exhibiting strong divergence. The directional-flow region within the sector toward +$k_x$, delineated by the grey dotted lines, is normalized at each *x* coordinate to eliminate polariton propagation losses along *x*.

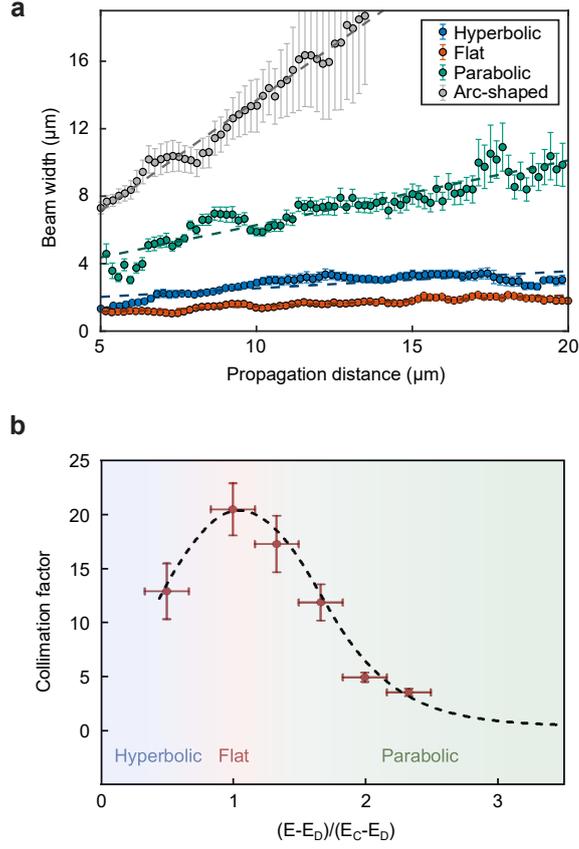

**Fig. 4| Reconfigurable directional polariton flow. a,** Quantitative analysis of polariton beam-width evolution for propagation along $+x$ across the hyperbolic, flat, and parabolic IFC regimes, compared with the pronounced divergence along $-y$ induced by arc-shaped IFCs. Error bars denote the 95% confidence intervals extracted from the beam-width fits at each propagation distance. Linear fits to the beam-width trends over propagation distance yield the corresponding lateral broadening rates. **b,** Collimation factor, given by the ratio of the lateral broadening rate of the arc-shaped IFC regime to that of each IFC regime, plotted as a function of the normalized energy offset $\xi = \frac{E-E_D}{E_c-E_D}$, where $E$, $E_D$, and $E_c$ denote the condensation energy, tilted Dirac-point energy, and canalization energy, respectively. The normalized energy offset $\xi$ delineates three regimes: a hyperbolic-IFC regime for $\xi < 1$, a flat-IFC (canalization) regime around $\xi \approx 1$, and a parabolic-IFC regime for $\xi > 1$. The black dashed curve represents the results given by theoretical simulations (see Methods). Vertical and horizontal error bars reflect uncertainties propagated from the lateral broadening rates and condensate linewidths, respectively.